\documentclass[prl,showpacs,twocolumn]{revtex4}
\usepackage{graphicx}
\begin{document}

\title{Scaling Behavior of Transverse Kinetic Energy Distributions in
Au+Au Collisions at $\sqrt{s_{\rm NN}}=200$ GeV}
\author{L. L. Zhu, H. Zheng and C. B. Yang}
\affiliation{Institute of Particle Physics, Hua-Zhong Normal
University, Wuhan 430079, People's Republic of China}

\begin{abstract}
With the experimental data from STAR on the centrality dependence
of transverse momentum $p_T$ spectra of pions and protons in Au+Au
collisions at $\sqrt{s_{NN}}=200\ {\rm GeV}$, we investigate the scaling
properties of transverse energy $E_T$ distributions at different centralities.
In the framework of cluster formation and decay mechanism for particle production,
the universal transverse energy distributions for pion and proton
can be described separately but not simultaneously.

\end{abstract}

\maketitle

\section{I. Introduction}

Recently, the scaling behaviors of the transverse momentum
distributions for particles produced in high energy collisions attract
more and more attentions. The search on the scaling
behaviors of particle spectrum is significantly important for
understanding the evolution of high energy heavy ion collisions
and particle production mechanisms, because the produced particle
distribution carries information about the dynamics of the system
and is one of the most important observables in high energy
collisions. In \cite{hy1,zhu1,hy2}, with the experimental data
from STAR, PHENIX and BRAHMS, we found there is a scaling behavior
of transverse momentum $p_T$ spectrum for pions. The scaling
function is independent of the colliding system, the colliding
energy, the centrality and (pseudo)rapidity. For protons and
anti-protons, there also exists a similar scaling behavior which
is independent of the centrality and rapidity in Au+Au collisions
at $\sqrt{s_{NN}}=200\ {\rm GeV}$ \cite{zhu2}. For strange
particles, such as $K$, $\Lambda$ and $\Xi$, the scaling behaviors
for their distributions are under consideration.

In high energy heavy ion collisions, a lot of the initial kinetic energy of
the bombarding nuclei is converted into those of the produced
particles, including longitudinal and transverse ones. Different
from momentum, kinetic energy is a scalar and is directly associated with
the temperature of the hot medium created in the collisions. For
different species of particles, the same momentum corresponds to
different kinetic energy because of mass effect. As well-known,
the kinetic energy, rather than momentum, of particles in a
thermalized system satisfies the Boltzmann distribution. Thus the
distribution of kinetic energy of particles produced in
ultra-relativistic heavy ion collisions is more effective in
revealing the thermal properties of the system. With above consideration in mind, one can ask
whether the transverse kinetic energy $E_T$ distribution has
similar scaling behavior. This is our motivation of investigating the scaling
properties of distributions of transverse kinetic energy $E_T$ of
particles. We hope the mass effect can be suppressed in the new scaling
function.

In this paper the scaling $E_T$ distribution of protons in the
mid-rapidity region is studied and compared with the scaling $p_T$
distribution. It is very essential to ask why the scaling
behaviors exist for different particles and what is the potential
universal dynamics. In \cite{zhu2} we found that the string overlap
mechanism can not explain data for pion and proton simultaneously.
In this paper we will consider another mechanism based on the parton percolation theory
\cite{cluster1,cluster2,cluster3}, and hope to get more
information about particle production mechanism in
nuclear collisions.

This paper is organized as follows: In Sec. II, we discuss relation
between distributions for $E_T$ and $p_T$, and the scaling
property of $E_T$ distribution of protons produced in Au+Au
collisions at $\sqrt{s_{NN}}=200\ {\rm GeV}$. Universal scaling functions
for pion and proton are given. In Sec. III
we discuss the universal scaling functions in the framework of cluster decay.
The conclusion is drawn in Sec. IV.

\section{II. The Scaling Behavior of Transverse Energy Distribution}
From the definition of transverse kinetic energy
$E_T\equiv m_T-m_0=\sqrt{p_T^2+m_0^2}-m_0$, with $m_0$ mass of
the particle, the invariant $E_T$ distribution is
\begin{equation}
\frac{d^2N}{E_TdE_Tdy}=(1+\frac{m_0}{E_T})\frac{d^2N}{p_Tdp_Tdy}\
.\label{ekpt}
\end{equation}
From the definition of $E_T$, one can see that when $m_0\rightarrow0$ or
$p_T\rightarrow\infty$, $E_T$ is approximately equal to $p_T$, and
\begin{equation}
\frac{d^2N}{E_TdE_Tdy}\approx\frac{d^2N}{p_Tdp_Tdy}\ .
\end{equation}
So it is feasible to only consider the scaling behavior of $p_T$
distribution in the range $p_T>0.5$ GeV/$c$ for pions, whose mass is
only $0.139\ {\rm GeV}$. Actually, the results obtained in \cite{hy1,zhu1,hy2} are
excellent. But if $m_0$ is large, as for kaons, protons and
anti-protons, etc, we must consider the mass effect and investigate the scaling
behaviors of $E_T$ distributions in the same range of $p_T$.

\begin{figure}[tbph]
\includegraphics[width=0.45\textwidth]{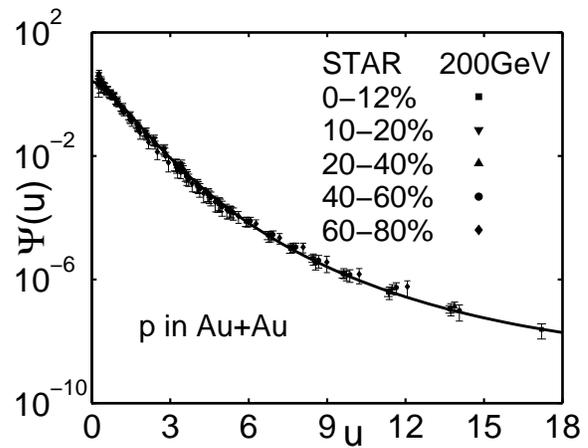}
\caption{Scaling behavior of the $E_T$ spectra for protons
produced at mid-rapidity in Au+Au collisions at RHIC. Feed-down
corrections are considered in the data. The solid curve is from
Eq. (\ref{protondis}). The data are taken from \cite{exp} after
calculation with Eq. (\ref{ekpt}).}\label{protonfd}
\end{figure}

In \cite{zhu2} we have summarized the method for searching the
scaling behavior of the particle's $p_T$ spectrum. Now we can use the
similar method to investigate the scaling behavior of the
transverse energy distribution of final state particles, with transverse
kinetic energy $E_T$ instead of transverse momentum $p_T$.
First, define a scaled variable
\begin{equation}
z=E_T/K\ ,
\end{equation}
and the scaled spectrum
\begin{equation}
\Phi(z)=A\left.\frac{d^2N}{2\pi E_TdE_Td\eta}\right|_{E_T=Kz}\ ,
\end{equation}
with $A$ and $K$ free parameters chosen to fit the scaled distributions
to the same curve. Of course values of $A$ and $K$ are different for distributions at different
centralities, given that for most central collisions they are set to be 1.
Then $\Phi(z)$ can be regarded as a parameterization of the $E_T$ distribution
in most central Au+Au collisions.
To obtain a universal scaling function for all centralities, independent of
the arbitrary in choosing values of $A$ and $K$ for the most central collisions,
we introduce another scaled variable
\begin{equation}
u=z/\langle z\rangle=E_T/\langle E_T\rangle\ ,\label{udef}
\end{equation}
and the normalized scaling function
\begin{equation}
\Psi(u)=\langle z\rangle^2\Phi(\langle z\rangle u)/\int_0^\infty
\Phi(z)zdz\ .\label{normal1}
\end{equation}
Here $\langle z\rangle$ is defined as
\begin{equation}
\langle z\rangle\equiv \int_0^\infty z\Phi(z)zdz/\int_0^\infty
\Phi(z)zdz \ ,
\end{equation}
and one can easily check that $\int du u\Psi(u)=1$ and
$\langle u\rangle=\int du u^2\Psi(u)=1$.
With above steps, one can investigate whether there exists a
scaling behavior of transverse energy distribution for protons
produced at mid-rapidity in Au+Au collisions at
$\sqrt{s_{NN}}=200\ {\rm GeV}$, which has a wide rang $p_T$
coverage \cite{exp}. As shown in Fig.
\ref{protonfd}, all data points for different centralities can be shifted
to the same curve within errorbar, and the corresponding values of parameters
$A$ and $K$ shown in TABLE I.  To parameterize the curve,
we define $v=\ln(1+z)$, and the scaling function is
\begin{equation}
\Phi(z)=18.604\exp(-0.6765v-6.48v^2+1.477v^3)\ .
\end{equation}
After normalization,
\begin{equation}
\Psi(u)=1.4846\exp(2.394v-5.79v^2+0.977v^3)\ , \label{protondis}
\end{equation}
with redefined $v=\ln(1+u)$.

\begin{table}
\def\tabcolsep{0.3cm}
\begin{tabular}{||c|c|c|c|c||}
\hline STAR & \multicolumn{2}{|c|}{$\pi$} & \multicolumn{2}{|c|}{$p$}\\
\hline  centrality & K & A & K & A\\
\hline  0-12\% & 0.4192 & 0.0038 & 0.5786 & 0.1440\\
\hline  10-20\% & 0.4293 & 0.0059 & 0.5769 & 0.1788\\
\hline  20-40\% & 0.4428 & 0.0114 & 0.5710 & 0.3022\\
\hline  40-60\% & 0.4535 & 0.0307 & 0.5635 & 0.6732\\
\hline  60-80\% & 0.4571 & 0.1025 & 0.5437 & 2.0985\\
\hline  40-80\% & 0.4536 & 0.0466 & & \\
\hline
\end{tabular}
\caption{Parameters for coalescing all data points to the same
curves in Figs. \ref{protonfd} for $p$ and \ref{pion} for $\pi^+$.}
\end{table}

\begin{figure}[tbph]
\includegraphics[width=0.45\textwidth]{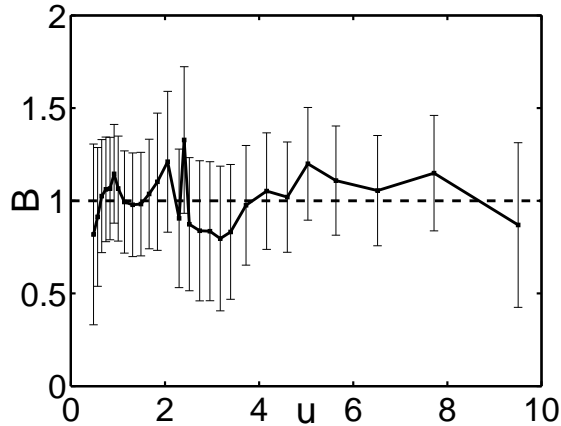}
\caption{The ratio $B$ of the experimental data to the scaling
$p_T$ fitted results Eq. (\ref{protonpt}), here $u=p_T/\langle
p_T\rangle$. Errorbars shown are calculated from those from experiment with
the definition of $B$.} \label{ratiopt}
\end{figure}

\begin{figure}[tbph]
\includegraphics[width=0.45\textwidth]{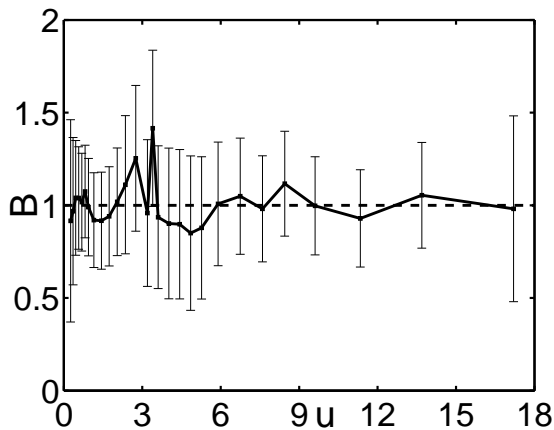}
\caption{The ratio $B$ of the experimental data to the scaling
$E_T$ fitted results Eq. (\ref{protondis}), here $u=E_T/\langle
E_T\rangle$.} \label{ratioet}
\end{figure}

In order to see the agreement between the scaling distribution and
the data in linear scale, a ratio $B$ can be defined as
\begin{center}
B=experimental data/fitted results.
\end{center}
First let us look at the agreement of the fit to the $p_T$ distribution.
The fitted normalized $p_T$ scaling function
of proton is obtained from \cite{zhu2}, which is given as follows,
\begin{equation}
\Psi(u)=0.064\exp(13.6v-16.67v^2+3.6v^3)\ , \label{protonpt}
\end{equation}
with $v=\ln(1+u)$ and $u=p_T/\langle p_T\rangle$. The result of $B$ for the
deviation of the scaling $p_T$ function from data in most central
collisions is shown in Fig. \ref{ratiopt}. They agree with each other within
experimental error.
Then one can investigate $B$ for the $E_T$ case. The result is shown in Fig.
\ref{ratioet}. As can be seen in Figs. \ref{ratiopt} and
\ref{ratioet}, in the middle region, both fits are good. But at the low
and high region, the agreement of the scaling $E_T$ distribution is better.
This indicates that the $E_T$ scaling can describe the experimental data better.
For data at other centralities, similar conclusion can be made.

\section{III. Scaling Transverse Kinetic Energy Distribution and Cluster Decay}

In \cite{hy1,hy2,zhu1,zhu2} and last section, we have found the scaling laws
of transverse momentum and kinetic energy spectra for pions, protons and anti-protons .
With those results, we should try to trace out the particle production mechanism. In
\cite{zhu2} we have shown that string picture for particle
production can not describe the scaling properties of $p_T$
distributions for both pions and protons simultaneously, because from the pion
and proton spectra one can get opposite changes of string overlap degree
from central to peripheral collisions.

 In the framework of parton percolation \cite{cluster1,cluster2,cluster3},
massive color-neutral clusters are formed in high energy collisions
from produced partons. In this picture all observed final state
particles are decay products of those clusters with different sizes.
Now we discuss whether such a picture can describe the transverse energy
distribution. One can use the inverse of transverse
kinetic energy squared, $x$, to describe the cluster size with
distribution denoted by $W(x)$. Here the transverse kinetic energy is used in place
of transverse momentum in \cite{cluster1,cluster2,cluster3}.
The normalized $W(x)$ has been
supposed to be a Gamma distribution
\begin{equation}
W(x)=\frac{\gamma}{\Gamma(k)}(\gamma x)^{k-1}\exp(-\gamma x)\
,\label{gammadis}
\end{equation}
with $\gamma$ and $k$ two parameters. To get the spectra of final
state particles one needs to know the cluster fragmentation
function $f(x,E_T)$ for each species of particles. Here $f(x,E_T)$ gives the
probability of producing a hadron with transverse kinetic energy $E_T$ from a cluster
of size $x$. We have no first-principle as
an instruction for the functional form of fragmentation function. We
expect that the fragmentation function is a function of the
fraction $z=E_T\sqrt{x}$ of transverse kinetic energy of the
produced particle relative to that of a cluster with size $x$. This fraction
is a close analogue of that for usual fragmentation functions from
hard partons to hadrons. For this reason, one can assume that the
fragmentation function for cluster decay takes the same functional
form as the usual ones, as used in  \cite{hy3} and references
therein,
\begin{equation}
f(z)=Dz^a(1-z)^b(1+cz^d)\ \label{frag}
\end{equation}
with $D$, $a$, $b$, $c$ and $d$ five parameters.

With the cluster size distribution and fragmentation function, the
$E_T$ distribution of a species of final state particles can be expressed as
\begin{equation}
\frac{dN}{E_TdE_T}=C\int_0^{1/E_T^2} dxf(E_T\sqrt{x})\times W(x)\
.\label{disek}
\end{equation}
In last equation $C$ is the normalization constant for the total number of
clusters formed before hadronization. In real fitting parameters $C$ and $D$
always appear as a product. So one can absorb $D$ into $C$.
The upper limit in the integration is given
from the consideration that hadron's fraction of transverse kinetic energy
cannot be larger than 1. The physics behind above formula is as
follows. The clusters are formed before their decay, so that
$W(x)$ is the same for all species of final state hadrons. On the
other hand, $f(z)$ is different for different hadrons, because
different hadrons come from different fragmentation channels of
clusters, but $f(z)$ should have no connection with the cluster
size $x$, thus is independent of the colliding system, centrality
and rapidity, namely universal for all collision processes. This is also a parallel
to the usual parton fragmentation functions which are independent of the processes
for the hard parton production. It is easy to
show that last equation guarantees scaling of the particle
distribution. Under the transformation $x\rightarrow\lambda x$,
$\gamma\rightarrow\gamma/ \lambda$ and $E_T\rightarrow
E_T/\sqrt{\lambda}$, both the two functions $W(x)$ and $f(z)$ are
all invariant, so the $E_T$ distribution is also
invariant within a normalization constant. This invariance means that
a change of the mean transverse kinetic energy can be equilibrated by a
change of the value of parameter $\lambda$ for the cluster distribution.
Thus it shares the essence with the well-known renormalization group transformation.
This invariance is the scaling we
are looking for.

\begin{figure}[tbph]
\includegraphics[width=0.45\textwidth]{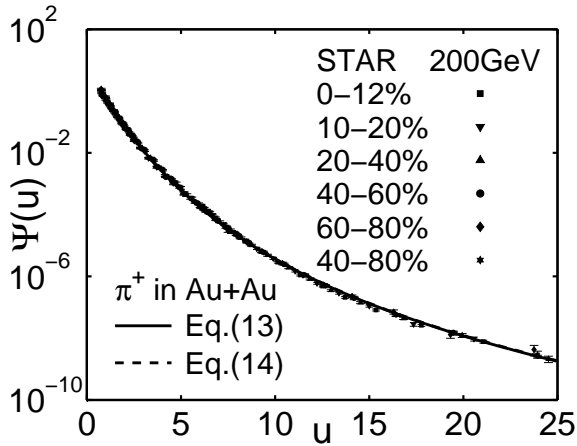}
\caption{Normalized scaling distribution for pions produced at
mid-rapidity in Au+Au collisions at RHIC with scaling variable
$u=p_T/\langle p_T\rangle$. Feed-down corrections also are
considered in the data. The solid curve is from Eq .(\ref{disek}),
and the dash curve is from Eq .(\ref{pionold}). The data is from
\cite{exp}.} \label{pion}
\end{figure}

\begin{figure}[tbph]
\includegraphics[width=0.45\textwidth]{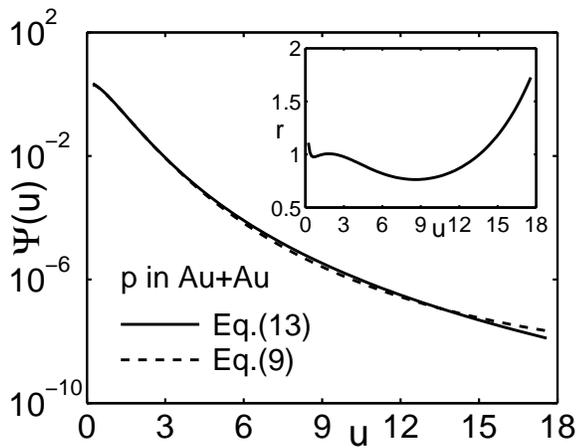}
\caption{Comparison between the normalized scaling distribution for protons produced at
mid-rapidity in Au+Au collisions at RHIC and the fitted result from the cluster
decay mechanism.} \label{proton}
\end{figure}

Now we show how to get the $E_T$ distributions of pions and
protons from the decay of clusters. Because pion is very light,
$E_T\simeq p_T$ in almost the whole range of observed transverse momentum, one only
need consider the scaling $p_T$ distribution. We use a recent set of data in
\cite{exp} for pion. Considering $p_T$ range in the data used now is much wider than that used
in \cite{hy1}, we fit the new data. All data points can be put to the same curve by
suitable $A$ and $K$, whose values are tabulated in TABLE I. The new parameterization for
the normalized scaling function is now
\begin{equation}
\Psi(u)=0.288\left(1+\frac{u^2}{7.868}\right)^{-4.3}(1+13.6e^{-1.9u}),\label{pionold}
\end{equation}
with $u=p_T/\langle p_T\rangle$. As shown in Fig .\ref{pion}, the agreement
between the scaling function and data is great. To describe the scaling
function for pion with the cluster mechanism, one can work with Eq. (\ref{disek})
and fit the obtained scaling function with 7 parameters, $C, \gamma$ and $k$
for cluster distributions, $a, b, c$ and $d$ for the fragmentation function
from a cluster to pion. The fitted parameters are tabulated in TABLE II.
For comparison, the fitted curve from the cluster decay is also plotted in Fig. \ref{pion}.
Two curves for the scaling function and the fitted result cannot be distinguished
in the plot in log scale. For protons, one should consider the scaling $E_T$ distribution,
as discussed in last section. To fit proton's scaling function, one cannot treat all
the three parameters for the cluster distribution as free parameters. As will be addressed
below, parameter $k$ must be the same in fitting the scaling functions for pion and proton.
The fitted parameters for proton case are also given in TABLE II, and the fitted curve
is shown in Fig. \ref{proton}. The normalized
scaling $E_T$ function Eq .(\ref{protondis}) is shown in this figure too. As
can be seen from Fig. \ref{proton}, the fit is also beautiful in a wide range of $E_T$.
The inset in Fig. \ref{proton} gives the ratio of the scaling
function for proton  to the fitted distribution from cluster decay.
It should be noted that the mean transverse momenta are different
for pion and proton in the collisions with given centrality. In obtaining the scaling
functions we try to scale the momentum (or transverse
kinetic energy) to a variable with unit mean. For this purpose,
the scaling factors for pions and protons must be different. The
scale in $p_T$ or $E_T$ can be translated into that of $x$ and finally to $\gamma$,
as discussed in the paragraph following Eq. (\ref{disek}). So the
values of $\gamma$ for pions and protons obtained in our fitting
are different. This difference in scaling factor also causes
different values of $C$, as shown in TABLE II. Another reason for the
difference of values of $C$ for pion and proton is that $C$ contains in fact
parameter $D$ as a factor which should be much smaller for proton
than for pion because of the low yield of proton relative to pion. Besides, the
distribution of clusters $W(x)$ is the same for all species of
final state particles, so one can figure out $k$ is the same for
pions and protons, because the scale change in $p_T$ or $E_T$ does not affect $k$.
Because the parameter $a$ for pion is about $-1$ while that for proton is only about 0.2,
the density of pion with low transverse kinetic energy (or momentum) is much higher
than for proton, in agreement with experimental fact that $p/\pi\sim 0$ as $p_T\to 0$.
 Now we can say the mechanism of cluster formation and decay can describe the
universal scaling behaviors of transverse energy distributions for
either pion or proton.

But can the cluster mechanism describe all those final state particle
spectra at the same time, as a true particle production mechanism should be able
to? The answer lies in the trends of change of the parameter $K$ for pion and proton from
central to peripheral collisions. $K$ increases for pions but decreases for protons.
The same results are also obtained in earlier works \cite{hy1,zhu1,hy2,zhu2} for the
$p_T$ distributions. Such trends are closely associated with the fact that the
pion spectrum is suppressed from intermediate to high $p_T$
in central Au+Au collisions while no suppression was observed for the proton
spectrum. In fact, from Eq. (\ref{udef}) one can see that the values of $K$
used in shifting data points to the normalized scaling distribution are the mean
values of transverse kinetic energy $\langle E_T\rangle$. With the cluster mechanism,
on the other hand, $\langle E_T\rangle$ can be calculated from Eq. (\ref{disek}) as
\begin{equation}
K=\langle E_T\rangle=\frac{\int_0^\infty dx W(x)/x^{3/2}}{\int_0^\infty dx W(x)/x}
\frac{\int_0^1 dz z^2f(z)}{\int_0^1 dz zf(z)}\ .
\end{equation}
As stated above, cluster distribution $W(x)$ is the same for different
species of particles, while cluster fragmentation functions $f(z)$ are
the same for a species of particle produced, independent of colliding centrality.
If we take a ratio between values of $K$ for different centralities, the ratio
must be the same for all particles, since the $f(z)$ terms cancelled in the  ratio.
This demands that values of $K$ for different species of particles must change
in the same way with centrality. This demand cannot be satisfied in the cluster
mechanism. This fact indicates that the
cluster formation and decay mechanism is not a universal one for the production of final
state pions, protons, and other particles in high energy collisions.
As shown in \cite{cluster3}, however, one can describe spectra of both pion
and (anti)proton with different values of the parameter $k$ in the cluster distribution.
Different values of $k$ implies that pions and (anti)protons originate from clusters with
different distributions.

Finally a brief discussion on the distributions at LHC energies can be addressed. In Pb+Pb
collisions at LHC energies $\langle E_T\rangle$ should be much larger than that at RHIC
energies. From the universal distribution discussed in this paper, the value of $\lambda$
in the cluster distribution must be much smaller, which means that the clusters formed
in those collisions at LHC energies must have mean transverse kinetic energy much larger
than that at RHIC energies.

\begin{table}
\def\tabcolsep{0.5cm}
\renewcommand{\arraystretch}{1.4}
\begin{tabular}{||c|c|c||}
\hline & $\pi$ & $p$\\
\hline  $C$ & 12.7766 & 0.3857\\
\hline  $\gamma$ & 203.7292 & 318.4096\\
\hline  $k$ & 4.3164 & 4.3164\\
\hline  $a$ & -1.0842 & 0.1890\\
\hline  $b$ & 10.0671 & 25.9110\\
\hline  $c$ & 201.0009 & 142.9827\\
\hline  $d$ & 0.0016 & 0.5406 \\
\hline
\end{tabular}
\caption{Values of parameters of the universal transverse energy
distributions Eq. (\ref{disek}) for pions and protons.}
\end{table}

\section{IV. Conclusion}
In this paper we discussed the scaling properties of the transverse kinetic
energy distributions for pions and protons in Au+Au collisions at 200 GeV.
We showed that the $E_T$ scaling behavior can describe experimental data
better than that for the $p_T$ case. Although the cluster formation and decay
mechanism can describe spectra of either pion or proton at different centralities,
it cannot give consistent description, with the same parameters in the cluster
distribution, for both species of particles at the same
time. Thus it can be excluded as a consistent particle production mechanism.
Other particle production mechanism must be in  effect for the scaling
behavior.

This work was supported in part by the National Natural
Science Foundation of China under Grant Nos. 10635020 and 10775057, by
the Ministry of Education of China under Grant No. 306022 and project IRT0624.

\end{document}